\documentclass[%
superscriptaddress,
reprint,
preprintnumbers,
amsmath,amssymb,aps,prl,longbibliography]{revtex4-1}

\usepackage{bm}
\usepackage[colorlinks=true,linktocpage=true,linkcolor=blue,citecolor=blue,urlcolor=blue,breaklinks]{hyperref} 

\usepackage{cleveref} 

\usepackage[utf8]{inputenc} 
\usepackage[T1]{fontenc} 
\usepackage{microtype}

\usepackage[mathscr]{euscript} 
\usepackage{relsize}

\newcommand{\half}{\frac{1}{2}}
\newcommand{\nn}{\nonumber}

\newcommand{\df}{\mathrm{d}}
\newcommand{\dow}{\partial} 

\usepackage{mathtools}

\usepackage{xparse}

\ExplSyntaxOn

\clist_map_inline:nn{A,B,C,D,E,F,G,H,I,J,K,L,M,N,O,P,Q,R,S,T,U,V,W,X,Y,Z}
{
  \expandafter\def\csname b#1\endcsname{\bm{#1}} 
  
  \expandafter\def\csname c#1\endcsname{\mathcal{#1}} 
  
  \expandafter\def\csname bc#1\endcsname{\bm{\mathcal{#1}}} 
  
  \expandafter\def\csname s#1\endcsname{{\mathsmaller{#1}}} 
  
  \expandafter\def\csname bb#1\endcsname{\mathbb{#1}} 
  
  \expandafter\def\csname rm#1\endcsname{\mathrm{#1}} 
  
  \expandafter\def\csname sc#1\endcsname{\mathscr{#1}} 
  
  \expandafter\def\csname sf#1\endcsname{\mathsf{#1}} 
  
  \expandafter\def\csname f#1\endcsname{\mathfrak{#1}} 
}

\ExplSyntaxOff

\DeclareMathOperator{\SU}{SU}

\newcommand{\lb}{\left (}
\newcommand{\rb}{\right )}

\newcommand\ext{\text{ext}}

\usepackage{blindtext}
\usepackage[bbgreekl]{mathbbol}

\begin{document} 

\title{Approximate symmetries, pseudo-Goldstones, and the second law of thermodynamics}

\author{Jay Armas}\email{j.armas@uva.nl}
\author{Akash Jain}\email{ajain@uva.nl}

\affiliation{Institute for Theoretical Physics, University of Amsterdam, 1090
  GL Amsterdam, The Netherlands}
\affiliation{Dutch Institute for Emergent Phenomena, 1090 GL Amsterdam, The Netherlands}

\author{Ruben Lier}\email{rubenl@pks.mpg.de}

\affiliation{Max Planck Institute for the Physics of Complex Systems, 01187 Dresden, Germany}
\affiliation{Wurzburg-Dresden Cluster of Excellence ct.qmat, 01187 Dresden,
  Germany} 


\begin{abstract}
  We propose a general hydrodynamic framework for systems with spontaneously
  broken approximate symmetries. The second law of thermodynamics naturally
  results in relaxation in the hydrodynamic equations, and enables us to derive
  a universal relation between damping and diffusion of pseudo-Goldstones. We
  discover entirely new physical effects sensitive to explicitly broken
  symmetries. We focus on systems with approximate U(1) and translation
  symmetries, with direct applications to pinned superfluids and charge density
  waves. We also comment on the implications for chiral perturbation theory.
\end{abstract} 

\pacs{Valid PACS appear here}

\maketitle

Symmetry has proved to be a powerful organisational tool in physics for
characterising and classifying phases of matter. Knowledge about the symmetries
of a physical system, and whether these are spontaneously broken by the
low-energy ground state, is often sufficient to develop an effective theory
describing its long-distance late-time behaviour. Symmetries are useful even
when they are only approximate. The canonical example of this is the extremely
successful effective theory for pions as Goldstones of spontaneously broken
SU(2) chiral symmetry. In this context, due to nonzero quark masses, the
symmetry is only approximate and the effective theory can be systematically
corrected to account for the pion mass.

In this paper, we draw general lessons about effective theories featuring
this \emph{pseudo-spontaneous} pattern of symmetry breaking. We are interested
in physical systems where an \emph{approximate global symmetry} is spontaneously
broken by the low-energy ground state, leading to a slightly massive
\emph{pseudo-Goldstone} field $\phi(x)$. The explicitly broken symmetry means
that the associated Noether charge conservation is weakly violated, giving rise
to physical effects such as relaxation, damping, and pinning.
Pseudo-spontaneous symmetry breaking is common across the phase space of matter,
due to inherent defects, inhomogeneities, and impurities in materials. Examples
include pinned crystals~\cite{PhysRevB.62.7553, PhysRevB.65.035312}, charge
density waves~\cite{PhysRevB.17.535, RevModPhys.60.1129, hartnolldelcretaz,
  Delacretaz:2019wzh}, pinned superfluids~\cite{Donos:2021pkk, Ammon:2021slb},
electrons in graphene \cite{Lucas:2017idv}, pinned
nematics~\cite{eringen2012microcontinuum, Delacretaz:2021qqu}, and pions in
chiral perturbation theory~\cite{PhysRevLett.88.202302, PhysRevD.66.076011,
  Grossi:2020ezz}, among many others.

In recent years, there have been several efforts towards developing hydrodynamic
techniques for systems with spontaneously broken approximate symmetries, aimed
at explaining experimental and holographic results; see
e.g.~\cite{hartnolldelcretaz, Grozdanov:2018fic, Amoretti:2018tzw,
  Delacretaz:2019wzh, Donos:2019txg, Donos:2019hpp, Amoretti:2019kuf,
  Ammon:2019wci, Andrade:2020hpu, Baggioli:2020edn, Donos:2021ueh,
  Amoretti:2021fch, Amoretti:2021lll}. A rigorous hydrodynamic framework for
these systems, however, is still missing in the literature. This is the primary
goal of this paper. We formulate a complete hydrodynamic theory for thermal
systems exhibiting pseudo-spontaneous symmetry breaking. 

The key accomplishment of our construction is to show that damping of
pseudo-Goldstones and charge (or momentum) relaxation follow from the second law
of thermodynamics. In particular, we derive the relation $\Omega=D_\phi k_0^2$
among the pseudo-Goldstone damping rate $\Omega$, attenuation $D_\phi$, and
correlation length $1/k_0$, first noted in holographic
models~\cite{Amoretti:2018tzw, Ammon:2019wci, Donos:2019hpp}. We emphasise that
our derivation only relies on the second law; see~\cite{Delacretaz:2021qqu} for
a derivation using the Schwinger-Keldysh effective field theory or locality of
the equations of motion~\footnote{See~\cite{Delacretaz:2021qqu} for critical
  comments on the derivation of \cite{Baggioli:2020nay, Baggioli:2020haa}.}.

Surprisingly, we find that dissipative effects also lead to many new transport
coefficients in the hydrodynamic theory that have not been identified in past
literature. The most notable of these is a coefficient $\lambda$ that enters the
Josephson equation for the pseudo-Goldstone at leading order as
$\dow_t\phi = \lambda\mu + \ldots$, where $\mu$ is the chemical potential
associated with the approximately conserved charge. In the context of pinned
crystals, this equation becomes $\dow_t\delta\phi^i = \lambda u^i + \ldots$,
where $\delta\phi^i$ is the displacement field and $u^i$ the fluid
velocity. While $\lambda=1$ when the symmetry is exact, we can generically have
$\lambda\neq 1$.  Another such coefficient $\lambda_T$ enters the entropy/heat
flux of pinned crystals at leading order as $s^i = (s+\lambda_T)u^i +
\ldots$. Physically, the coefficients of this type result in a modification of
the speed of sound dependent on the strength of explicit symmetry breaking.

\vspace{1em}
\noindent
\emph{Pinned simple diffusion.}---To highlight the striking features of our
construction, we start with a simple toy model where the only conserved quantity
is a scalar charge density $n$. This could be mass, energy, number of particles,
electromagnetic charge, etc. We later generalise this construction to
approximately conserved momenta resulting in a theory of pinned crystals.

According to Noether's theorem, we expect that the system exhibits an associated
global symmetry. However, this symmetry could be spontaneously broken in the
low-energy ground state, giving rise to a Goldstone field $\phi(x)$. The
equilibrium configurations of the system can be described by the free energy
$F = \int\df^dx\,(\mathcal{F}(\phi) - K_\ext \phi)$, where $K_\ext$ is an
external source coupled to $\phi$. The free energy density $\mathcal{F}(\phi)$
obeys a Goldstone shift symmetry $\phi(x)\to\phi(x) - \Lambda$. This forbids a
mass term like $\mathcal{F} \sim \half m^2 \phi^2$ in the free energy density,
rendering the Goldstone massless.

The situation is qualitatively different when the said conservation is only
approximate, because the shift symmetry need not be respected. In practice,
however, we find it convenient to artificially restore the symmetry by coupling
the system to a background field $\Phi(x)$ that also shifts as
$\Phi(x)\to \Phi(x) - \Lambda$. The background field explicitly breaks the
global symmetry by picking out a preferred phase $\Phi$. We can now write a mass
term in $\mathcal{F}$
\begin{equation}
  \mathcal{F} = -p + \half f_s\,\dow_i\phi\dow^i\phi
  + \half\ell^2 m^2(\phi-\Phi)^2,
  \label{eq:free-U1}
\end{equation}
where $\ell$ is a bookkeeping parameter that controls the strength of explicit
symmetry breaking. The free energy $F$ with \eqref{eq:free-U1} can be understood
as a generalised Ginzburg-Landau model that accounts for explicit symmetry
breaking with an arbitrary source $\Phi$; see e.g.~\cite{2015GL}. The mass term
can be thought of as an ``elastic potential'' that tends to align the phase
$\phi$ with the background phase $\Phi$. Varying~\eqref{eq:free-U1} results in a
configuration equation
\begin{subequations}
  \begin{gather}
    f_s\!\lb\dow_i\dow^i\phi - k_0^2\phi\rb + m^2\ell^2\Phi + K_\ext = 0,
    \label{eq:config-static-U1}
  \end{gather}
  where $k_0=\ell m/\sqrt{f_s}$ is the finite inverse correlation length for
  $\phi$, demoting it to a massive \emph{pseudo-Goldstone}. Typically, this
  massive field can be integrated out from the long-wavelength effective theory.
  However, if the symmetry is still approximately preserved, i.e. $\ell$ is
  sufficiently small, the pseudo-Goldstone can still affect the long-wavelength
  spectrum. Using the Noether procedure, the static Ward identity
  \begin{equation}
    \dow_i j^i = K_\ext + m^2\ell^2(\phi-\Phi),
    \label{eq:static-conservation-U1}
  \end{equation}
\end{subequations}
follows from \eqref{eq:free-U1}, where $j^i = -f_s\dow^i\phi$ is the charge
flux.  As expected, the mass term results in a violation of charge conservation
even in the absence of external sources.

When we leave thermal equilibrium, we can no longer start with a free-energy and
must rely on the framework of hydrodynamics to proceed. Firstly, we have a
\emph{Josephson equation} giving dynamics to $\phi$ which, generalising
\eqref{eq:config-static-U1}, we take to have the schematic form
\begin{subequations}
  \begin{equation} \label{eq:gengold}
    K+K_\ext=0.
  \end{equation}
  We also have a conservation equation, generalising
  \eqref{eq:static-conservation-U1}, describing the dynamics of charge density
  $n$
  \begin{equation} \label{eq:cU1}
    \dow_t n + \dow_i j^i = -K -\ell L,
  \end{equation}
  where $L$ is some operator causing explicit symmetry breaking. We will also
  need a new energy conservation equation implementing the first law of
  thermodynamics
  \begin{equation}
    \dow_t\epsilon + \dow_i\epsilon^i
    = - K\dow_t\phi
    - \ell L \dow_t\Phi,
    \label{eq:energy-U1}
  \end{equation}
  where $\epsilon,\epsilon^i$ are the energy density and flux respectively. To
  complete these equations, we must give a set of constitutive relations for
  $j^i,\epsilon^i,K,L$ in terms of $n,\epsilon,\phi,\Phi$, arranged
  order-by-order in gradients.
  \label{eq:U1-equations}
\end{subequations}

We implement the gradient counting scheme where
$\phi\sim\mathcal{O}(\dow^{-1})$, making its gradients $\mathcal{O}(\dow^0)$;
see~\cite{Armas:2019sbe, Armas:2020bmo}. We ascribe the scaling
$\mathcal{O}(\dow)$ to the symmetry breaking parameter $\ell$ and require that
all dependence on the background phase $\Phi$ must appear via
$\psi=\ell(\phi-\Phi)\sim\mathcal{O}(\dow^0)$. This ensures that setting
$\ell=0$ restores the symmetry.

The most important ingredient in a hydrodynamic theory is the local second law
of thermodynamics. It necessitates the existence of an entropy density $s^t$ and
flux $s^i$ such that
\begin{equation}
  \dow_t s^t +\dow_i s^i \geq 0,
\end{equation}
is satisfied for \emph{every} solution of the conservation equations;
see~\cite{Jain:2018jxj}. Despite being an inequality, this requirement is
extremely powerful and is known to give strong constraints on the constitutive
relations~\cite{landau2013fluid}. At leading order in gradients, we simply have
$s^t = s(\epsilon,n,\dow_i\phi\dow^i\phi,\psi)$.  We can define the temperature
$T$, chemical potential $\mu$, ``superfluid density'' $f_s$, pseudo-Goldstone
mass parameter $m$, and the grand-canonical free-energy density $\mathcal{F}$
via
\begin{align}
  T\df s
  &= \df\epsilon - \mu\, \df n
    - \half f_s \df (\dow_i\phi\dow^i\phi)
    - m^2\psi\df\psi
    , \nn\\
  \mathcal{F}
  &= \epsilon - Ts^t - \mu n.
    \label{eq:U1-thermo}
\end{align}
The entropy density will, in general, admit gradient corrections. We consider
these in the appendix.

For clarity, let us assume the dynamics to be isothermal, i.e. $T=T_0$, so that
energy conservation decouples from the charge conservation and Josephson
equations. In this case, the second law constraints result in
\begin{align}
  j^i
  &= - f_s\dow^i\phi - \sigma_n\dow^i\mu, \nn\\
  K
  &= \dow_i \big( f_s\dow^i\phi \big)
    - \ell m^2 \psi 
    - (\sigma_\phi+\ell\sigma_\times)(\dow_t\phi-\mu)
    + \sigma_\times\dow_t\psi, \nn\\
    L
  &= m^2\psi + \sigma_\Phi\dow_t\psi
    - (\sigma_\times+\ell\sigma_\Phi)(\dow_t\phi-\mu).
\end{align}
Setting $\mu=\mu_0$ and $\dow_t\phi=\dow_t\Phi=\mu_0$, at leading order in
gradients, we recover the equilibrium version of these equations derived from
\eqref{eq:free-U1}. We also find four dissipative coefficients
$\sigma_n,\sigma_\phi,\sigma_\Phi,\sigma_\times$ that satisfy the inequality
relations $\sigma_n,\sigma_\phi\geq 0$ and
$\sigma_\Phi \geq \sigma_\times^2/\sigma_\phi$. We provide a detailed derivation
in the appendix.

\vspace{1em}
\noindent
\emph{Linearised fluctuations.}---To highlight the physical implication of this
model, we set $\Phi=\mu_0t$, $K_\ext=0$, and linearly expand the equations
around the solution $\mu=\mu_0$, $\phi=\mu_0t$. We find
\begin{subequations}
  \begin{align}
    j^i
    &= - f_s\dow^i\delta\phi
      - \sigma_n\dow^i\delta\mu, \nn\\
    \dow_t\delta\phi
    &= \lambda\,\delta\mu
      - \Omega\,\delta\phi
      + D_\phi\dow_i\dow^i\delta\phi, \nn\\
    \ell L
    &= \frac{\chi}{\lambda}\omega_0^2\delta\phi
      + \Gamma\delta\mu
      + (1-\lambda)f_s \dow_i\dow^i\delta\phi,
  \end{align}
  where we have defined the susceptibility $\chi$, pinning frequency $\omega_0$,
  pseudo-Goldstone attenuation constant $D_\phi$, damping constant $\Omega$,
  charge relaxation coefficient $\Gamma$, and a new coefficient $\lambda$ as
  \begin{gather}
    \chi = \frac{\dow n}{\dow\mu}, \quad
    \omega_0^2 = \frac{\lambda^2\ell^2m^2}{\chi}, \quad
    D_\phi= \frac{f_s}{\sigma_\phi}, \quad
    \Omega = \frac{\ell^2 m^2}{\sigma_\phi}, \nn\\
    \Gamma = \frac{\ell^2}{\chi} \lb\sigma_\Phi -
    \frac{\sigma_\times^2}{\sigma_\phi}\rb, \quad
    \lambda = 1+ \frac{\ell\sigma_\times}{\sigma_\phi}.
  \end{gather}%
  \label{eq:U1-constraints}%
\end{subequations}
Solving the equations and assuming $\ell\sim\mathcal{O}(k)$, we find a damped
sound mode with dispersion relations
\begin{align}
  \omega
  &= \pm \sqrt{\omega_0^2+v_s^2k^2} - \frac{i}{2}\lb k^2
    (D_n+D_\phi)+\Gamma+\Omega \rb,
    \label{eq:sound-U1}
\end{align}
where $v_s^2 = \lambda^2f_s/\chi$ and $D_n = \sigma_n/\chi$. The second law
inequality constraints imply that $D_n,D_\phi,\Omega,\Gamma\geq 0$, ensuring
that the sound pole remains in the lower-half plane.

From \eqref{eq:U1-constraints}, it is possible to make a few interesting
observations.  For instance, we have proved the damping-attenuation relation
\begin{equation}
  \Omega = D_\phi k_0^2,
  \label{eq:damping-relation-U1}
\end{equation}
where $k_0 = \omega_0/v_s$~\footnote{This relation was recently derived
  in~\cite{Delacretaz:2021qqu} by invoking the locality of constitutive
  relations. However, our derivation merely follows as a consequence of the
  second law of thermodynamics.}. We also see that our model naturally gives rise
to charge relaxation $\Gamma$, without needing to introduce it by
hand. Interestingly, we also find a new coefficient $\lambda\neq 1$ appearing in
front of the $\mu$ term in the Josephson equation. It affects the dispersion
relations by modifying the speed of sound in the presence of small explicit
symmetry breaking, and has not appeared in the literature before.

It might be tempting to try and absorb the new coefficient $\lambda$ via some
rescaling of fields and coefficients. However, it is possible to find a Kubo formula
for $\lambda$ using the static response functions
\begin{equation} \label{eq:lambdaK}
  \lambda^2 = -v_s^2 \frac{G^R_{nn}(\omega=0,k=0)}{G^R_{j^xj^x}(\omega=0,k=0)},
\end{equation}
where the speed of sound $v_s$ can be read off using the singularity structure
of the dispersion relations \eqref{eq:sound-U1}. This gives $\lambda$ a concrete
physical meaning. We have included further details regarding the physical
implications of $\lambda$ in the appendix.

The discussion can be extended to account for temperature fluctuations and
conserved momenta, leading to a theory of explicitly broken superfluids. This
theory was recently considered in the holographic context
in~\cite{Ammon:2021slb}. It will be interesting to revisit their results in the
view of our new $\lambda$ coefficient, along with other similar coefficients
that can appear in the energy flux and stress tensor. We will discuss this in
more detail in another publication.

\vspace{1em}
\noindent
\emph{Pinned viscoelastic crystals.}---The hydrodynamic theory for pinned
crystals can be constructed similarly to the U(1) case. In $d$ spatial dimensions,
a static configuration of a crystal can be described by the spatial distribution
of its lattice sites $\phi^{I=1,\ldots,d}(x)$, called the \emph{crystal
  fields}. We can define the \emph{strain tensor} as
$u_{IJ} = \half(h_{IJ}-\delta_{IJ}/\alpha_0^2)$, where $h_{IJ}$ is the inverse
of $h^{IJ}=\dow^i\phi^I\dow_i\phi^J$ and $\alpha_0$ is a constant parametrising
the ``inverse lattice spacing'' of the crystal. We can always rescale the fields
to set $\alpha_0=1$.

When the crystal is homogeneous, the free-energy density $\mathcal{F}$ obeys a
global spatial shift symmetry $\phi^I(x)\to\phi^I(x)+a^I$, and $\phi^I$ can be
understood as Goldstones of spontaneously broken translations. However, when the
crystal has slight inhomogeneities, possibly due to defects or impurities, this
shift symmetry can be violated, and $\phi^I$ become pseudo-Goldstones of
approximate translation symmetry. Analogous to the U(1) case, we
artificially restore the symmetry by introducing a set of background fields
$\Phi^I(x)$, also shifting as $\Phi^I(x)\to\Phi^I(x)+a^I$. In the present case,
$\Phi^I(x)$ can be interpreted as describing the spatial configuration of a
fixed background lattice coupled to our physical crystal of interest. This
allows us to introduce a mass term in the free-energy density
\begin{equation}
  \mathcal{F} = -p+ \half (B-{\textstyle\frac{2}{d}}G) (u^I_{~I})^2
  + G\, u^{IJ} u_{IJ}
  +  \frac{m^2}{2} \psi_I\psi^I,
  \label{eq:crystal-EOS}
\end{equation}
where $\psi^I=\ell(\phi^I-\Phi^I)$ is the \emph{misalignment tensor}. $B$ and
$G$ are bulk and shear moduli respectively. $I,J,\ldots$ indices are
raised/lowered using $h^{IJ}$, $h_{IJ}$.

To describe the dynamical evolution of this system, we need to formulate the
theory of pinned viscoelastic hydrodynamics following the construction
of~\cite{Armas:2019sbe, Armas:2020bmo}. Firstly, analogous to
\eqref{eq:gengold}, we have a set of Josephson equations for the crystal fields
\begin{subequations}
  \begin{equation} \label{eq:JoPinned}
    K_I + K_I^\ext = 0,
  \end{equation}
  where $K_I$ is an unknown operator and $K_I^\ext$ are sources coupled to
  $\phi^I$.  Assuming the crystal to exhibit Galilean symmetry, we also have
  momentum conservation and continuity equations
  \begin{align}
    \dow_t\pi^i + \dow_j\tau^{ij}
    &= K_I \dow^i\phi^I + \ell L_I \dow^i\Phi^I, \nn\\
    \dow_t\rho + \dow_i\pi^i
    &= 0,
      \label{mom-conservation-crystal}
  \end{align}
  where $\pi^i$ is the momentum density, $\tau^{ij}$ the stress tensor, $\rho$ the mass
  density, and $L_I$ an operator causing explicitly broken translations. These
  have to be supplemented with the energy conservation equation arising from the
  first law of thermodynamics
  \begin{equation}
    \dow_t\epsilon + \dow_i\epsilon^i
    = - K_I\dow_t\phi^I
    - \ell L_I \dow_t\Phi^I.
  \end{equation}
  We can now proceed and derive a set of constitutive relations
  for $\tau^{ij},\epsilon^i,K_I,L_I$ in terms of
  $\pi^i,\epsilon,\rho,\phi^I,\Phi^I$, arranged in a gradient expansion, and
  obtain constraints due to the second law of thermodynamics.
\end{subequations}

Imposing Galilean invariance, at leading order in gradients, the entropy density
is given by $s^t = s(\varepsilon,\rho,h^{IJ},\psi^I)$, where
$\varepsilon = \epsilon-{\textstyle\half}\rho\vec u^2$ is the Galilean-invariant
``internal energy density'' and $u^i=\pi^i/\rho$ is the fluid velocity. We can
define the temperature $T$, chemical potential $\mu$, elastic stress tensor
$r_{IJ}$, pseudo-Goldstone mass $m$, and free-energy $\mathcal{F}$ via
\begin{align}
  T\df s
  &= \df\varepsilon
    - \mu\, \df\rho
    + \half r_{IJ}\df h^{IJ} - m^2\psi_I\df\psi^I, \nn\\
  \mathcal{F}
  &= \varepsilon -Ts^t - \mu\rho.
    \label{eq:thermo-crystal}
\end{align}
The entropy density can also admit first order gradient corrections, which we
consider in detail in the appendix.

Similarly to the U(1) case, restricting to an isothermal regime, i.e. $T=T_0$,
energy conservation decouples from the rest of the system, and we obtain the
allowed set of constitutive relations
\begin{align}
  \tau^{ij}
  &= \rho u^i u^j
    - \mathcal{F} \delta^{ij}
    - r_{IJ} \dow^i\phi^I\dow^j\phi^J
    - 2\eta\dow^{\langle i} u^{j\rangle}
    - \zeta\dow_k u^k\delta^{ij}, \nn\\
  K_I
  &= -\dow_i\big( r_{IJ}\dow^i\phi^J\big)
    - \ell m^2\psi_I \nn\\
  &\qquad
    - (\sigma_\phi+\ell\sigma_\times)h_{IJ}\frac{\df\phi^J}{\df t}
    + \sigma_\times h_{IJ} \frac{\df\psi^J}{\df t}, \nn\\
  L_I
  &= m^2\psi_I
    + \sigma_\Phi h_{IJ}\frac{\df\psi^J}{\df t}
    - (\sigma_\times-\ell\sigma_\Phi) h_{IJ}\frac{\df\phi^J}{\df t},
\end{align}
where $\df/\df t = \dow_t + u^i\dow_i$ and angular brackets around indices
denote a symmetric-traceless combination. The five dissipative coefficients
$\eta,\zeta,\sigma_\phi,\sigma_\Phi,\sigma_\times$ follow the inequalities
$\eta,\zeta,\sigma_\phi\geq0$ and $\sigma_\Phi\geq
\sigma_\times^2/\sigma_\phi$. A detailed derivation relaxing the isothermal
assumption appears in the appendix.

\vspace{1em}
\noindent
\emph{Linear pinned crystals.}---In the small strain regime, the equation of
state of the crystal can be written as \eqref{eq:crystal-EOS}, except that
$\alpha_0$ in $u_{IJ}$ should now be promoted to $\alpha(T,\mu)$. We can still
set $\alpha(T_0,\mu_0)=1$ by rescaling the fields, but its thermodynamic
derivatives are generically nontrivial. Setting $\Phi^I = x^I$ and
$K^\ext_I = 0$, and expanding around $\phi^I=\delta^I_i(x^i-\delta\phi^i)$,
$\mu=\mu_0$, $u^i=0$ we can obtain
\begin{subequations}
  \begin{align}
    \tau^{ij}
    &= \lb p + B\alpha_m\delta\mu\rb \delta^{ij}
      - 2\eta \dow^{\langle i} u^{j\rangle}
      - \zeta\dow_k u^k\delta^{ij} \nn\\
    &\quad
      - B \dow_k \delta\phi^k\delta^{ij}
      - 2G \dow^{\langle i} \delta\phi^{j\rangle}, \nn\\[0.2em]
    \dow_t\delta\phi^i
    &= \lambda u^i - \Omega\delta\phi^i
      + \gamma_m\dow^i\mu \nn\\
    &\qquad
      + 2D_\phi^\perp\dow_k\dow^{[k} \delta\phi^{i]}
      + D_\phi^\|\dow^i\dow_k \delta\phi^k, \nn\\[0.2em]
    \ell L^i
    &= -\frac{\rho}{\lambda}\omega_0^2\delta\phi^i
      - \Gamma \pi^i
      - (\lambda-1)B\alpha_m\dow^i\mu \nn\\
    &\quad
      + (\lambda-1) \lb B\dow^i\dow_k \delta\phi^k
      + 2G\dow_k\dow^{\langle j} \delta\phi^{i\rangle} \rb.
  \end{align}
  We have defined pinning frequency $\omega_0$, mass expansion coefficient
  $\alpha_m$, pseudo-Goldstone attenuation constants $D^{\perp,\|}_\phi$,
  damping constant $\Omega$, momentum relaxation $\Gamma$, and coefficients
  $\lambda,\gamma_m$ as
  \begin{gather}
    \omega_0^2 = \frac{\lambda^2\ell^2m^2}{\rho},
    \quad \alpha_m = -\frac{d}{\alpha}\frac{\dow\alpha}{\dow\mu}, \quad
    \gamma_m = -\frac{B\alpha_m}{\sigma_\phi}, \nn\\
    D_\phi^\perp = \frac{G}{\sigma_\phi}, \quad D_\phi^\| = \frac{B+
      2\frac{d-1}{d}G}{\sigma_\phi}, \quad
    \Omega = \frac{\ell^2 m^2}{\sigma_\phi }, \nn\\
    \Gamma = \frac{\ell^2}{\rho} \lb \sigma_\Phi -
    \frac{\sigma_\times^2}{\sigma_\phi }\rb, \quad \lambda =
    1+\frac{\ell\sigma_\times}{\sigma_\phi}.
  \end{gather}
  \label{eq:crystal-linear}%
\end{subequations}
Looking at the mode spectrum, we obtain a damped sound mode in the longitudinal
and transverse sectors similar to \eqref{eq:sound-U1}. We also find a crystal
diffusion mode in the longitudinal sector. These expressions are given in the
appendix. Lifting the isothermal assumption leads to an energy diffusion mode
coupled with crystal diffusion; see e.g.~\cite{Armas:2020bmo}.

Analogously to the U(1) case, using~\eqref{eq:crystal-linear} we recover the
damping-attenuation relation from~\cite{Amoretti:2018tzw, Ammon:2019wci,
  Donos:2019hpp}
\begin{equation}
  \Omega = D_\phi^\perp k_0^2,
  \label{eq:damping-relation-crystal}
\end{equation}
where $k_0 = \omega_0/v_\perp$ with $v_\perp^2 = \lambda^2G/\rho$.
The
momentum relaxation $\Gamma$ also arises naturally in our model, along with the
coefficient $\lambda$ affecting the Josephson equation. Upon including thermal
fluctuations, we find another damping-attenuation relation similar to
\eqref{eq:damping-relation-crystal} in the energy flux.  We also find a new
pinning-sensitive coefficient $\lambda_T$ in the energy flux; see the appendix
for more details.

\vspace{1em}
\noindent
\emph{Discussion.}---In this paper we introduced a general hydrodynamic
framework for dissipative systems with spontaneously broken approximate
symmetries. Our construction builds upon the technology of forced fluid dynamics
from~\cite{Bhattacharyya:2008ji, Armas:2016mes}, by systematically coupling the
hydrodynamic equations to pseudo-Goldstone fields $\phi(x)$ and fixed background
phase fields $\Phi(x)$, responsible for spontaneously and explicitly breaking
the symmetries respectively~\footnote{In the context of holography, forced fluid
  dynamics and explicit symmetry breaking have been studied in multiple works,
  see e.g.~\cite{Andrade:2013gsa, Andrade:2014xca, Andrade:2015iyf,
    Blake:2015epa, Andrade:2015hpa}}. We illustrated how the interplay between
the two field ingredients gives rise to physical effects such as damping,
pinning, and relaxation. In particular, we showed that the elusive relation
between the damping $\Omega$ and attenuation $D_\phi$ of pseudo-Goldstones
follows simply by imposing the second law of thermodynamics in the presence of
background fields $\Phi(x)$. The second law also requires the relaxation
coefficient $\Gamma$ to be non-negative.

In addition to providing a rigorous mathematical language for systems with
pseudo-spontaneously broken symmetries, we also found entirely new physical
effects that have not been discussed in previous literature. Namely, we
discovered new transport coefficients sensitive to the explicit nature of
symmetry breaking that modify the hydrodynamic and Josephson equations at the
thermodynamic level. These coefficients result in a modification of the speed of
the damped sound mode and affect the hydrodynamic correlators in a non-trivial
way.

We primarily focused on systems with approximate U(1) or approximate spatial
translation symmetry. However, the framework developed here is equally relevant
for other physical situations exhibiting a pseudo-spontaneous pattern of
symmetry breaking. For instance, a hydrodynamic theory for pions recently
appeared in~\cite{Grossi:2020ezz}, featuring a pseudo-spontaneously broken SU(2)
chiral symmetry~\cite{PhysRevLett.88.202302, PhysRevD.66.076011,
  Jain:2016rlz}. In particular,~\cite{Grossi:2020ezz} noted that the
damping-attenuation relation \eqref{eq:damping-relation-U1} for pions follows
from the second law of thermodynamics. However, their analysis does not include
additional pinning-sensitive coefficients such as $\lambda$. It is
straight-forward to generalise the U(1) case analysed here to a SU(2) pion field
$\phi^a$ coupled to a fixed background SU(2) phase $\Phi^a$, where $a,b,\ldots$
denote $\SU(2)$ Lie algebra indices.
The linearised Josephson equation will take the schematic form
\begin{equation}
  \dow_t\delta\phi^a
  = \lambda^a{}_b\,\delta \mu^b
  - \Omega^a{}_b\, \delta\phi^b
  + D_\phi{}^a{}_b\, \dow_i\dow^i\delta\phi^b,
\end{equation}
with the damping-attenuation relation $\Omega^a{}_b = D_\phi{}^a{}_b\,
k_0^2$. The coefficient $\lambda^a{}_b$ is equal to $\delta^a_b$ in the absence
of explicit symmetry breaking, but can acquire corrections when the symmetry is
weakly broken, modifying the mode spectrum of chiral perturbation theory.

Holographic models with pseudo-spontaneous pattern of symmetry breaking have
been discussed in multiple works; see e.g. \cite{Amoretti:2018tzw,
  Donos:2019txg, Donos:2019hpp, Amoretti:2019kuf, Andrade:2020hpu,
  Baggioli:2020edn, Donos:2021ueh, Amoretti:2021fch, Amoretti:2021lll,
  Ammon:2019wci}. It would be interesting to develop the relativistic versions
of the hydrodynamic theories formulated in this paper and revisit their
holographic applications in light of the new transport coefficients that we have
identified. We leave this direction for future work.

\vspace{1em}

\begin{acknowledgments}
  JA and AJ are partly supported by the Netherlands Organization for Scientific
  Research (NWO) and by the Dutch Institute for Emergent Phenomena (DIEP)
  cluster at the University of Amsterdam. AJ is funded by the European Union’s
  Horizon 2020 research and innovation programme under the Marie
  Sk\l{}odowska-Curie grant agreement NonEqbSK No.~101027527.
\end{acknowledgments}

\bibliography{pinnedEntropy} 

\clearpage

\appendix
\renewcommand\thefigure{A\arabic{figure}}    
\setcounter{figure}{0} 
\renewcommand{\theequation}{A\arabic{equation}}
\setcounter{equation}{0}

{\center{\large\bfseries Supplementary Material}\par}

\section{Details of pinned simple diffusion}

In this appendix we provide details of the second law analysis in the pinned
simple diffusion model. We also turn on a background gauge field $A_t$, $A_i$
throughout the discussion to facilitate the computation of hydrodynamic
correlation functions. In addition, we do not assume the system to be isothermal
as in the bulk of the paper.

\vspace{1em}
\noindent 
\emph{Second law constraints}.---We can define the gauge covariant derivatives
of $\phi$ and $\Phi$ as
\begin{gather}
  \xi_t = \dow_t\phi+A_t, \qquad
  \xi_i = \dow_i\phi+A_i, \nn\\
  \Xi_t = \dow_t\Phi+A_t, \qquad
  \Xi_i = \dow_i\Phi+A_i.
\end{gather}
In the presence of background gauge fields, the Josephson and charge
conservation eqs.~\eqref{eq:gengold}-\eqref{eq:cU1} remain unchanged, but the
energy conservation \eqref{eq:energy-U1} generalises to
\begin{equation}
  \dow_t\epsilon + \dow_i\epsilon^i
  = E_ij^i - K \xi_t
  - \ell L \Xi_t,
\end{equation}
where $E_i = \dow_i A_t - \dow_t A_i$. Let us parametrise the entropy density as
\begin{equation}
  s^t = s + \mathcal{S},
\end{equation}
where $\mathcal{S}$ are possible gradient corrections. Using the thermodynamic
relations \eqref{eq:U1-thermo} and the conservation equations, we find
\begin{align}
  \dow_t s^t + \dow_is^i
  &= -  \frac{1}{T^2}\mathcal{E}^i\dow_i T
    - \mathcal{J}^i \lb \dow_i \frac{\mu}{T} - \frac{E_i}{T}\rb \nn\\
  &\hspace{-4em}
    - \frac1T\mathcal{K}(\xi_t-\mu)
    - \frac{\ell}{T} \mathcal{L}(\Xi_t-\mu)
    + \dow_t \mathcal{S} + \dow_i\mathcal{S}^i,
    \label{eq:adiabaticity-U1}
\end{align}
where we have identified the constitutive relations
\begin{align}
  \epsilon^i
  &= -f_s \xi^i\xi_t + \mathcal{E}^i, \nn\\
  j^i
  &= -f_s\xi^i + \mathcal{J}^i, \nn\\
  K
  &= \dow_i (f_s\xi^i) - \ell m^2 \psi + \mathcal{K}, \nn\\
  L
  &= m^2\psi + \mathcal{L}, \nn\\
  s^i
  &= \frac1T \mathcal{E}^i
    - \frac{\mu}{T} \mathcal{J}^i
    + \mathcal{S}^i.
\end{align}

The right-hand side of \eqref{eq:adiabaticity-U1} is required to be a positive
semi-definite quadratic form. Truncating at first order in gradients, there are two
kinds of solutions to \eqref{eq:adiabaticity-U1}. First, we have the
``non-hydrostatic sector'', where $\mathcal{S}$, $\mathcal{S}^i$ are identically
zero and we simply have
\begin{equation}
  \begin{pmatrix}
    \frac1T \mathcal{E}^i_{\text{nhs}} \\
    \mathcal{J}^i_{\text{nhs}} \\
    \mathcal{K}_{\text{nhs}} \\
    \mathcal{L}_{\text{nhs}}
  \end{pmatrix}
  = -
  \begin{pmatrix}
    \frac1T\kappa^{ij} & \gamma^{ij} & \gamma_{\epsilon\phi}^i & \gamma_{\epsilon\Phi}^i \\
    \gamma'^{ij} & \sigma_n^{ij} & \gamma_{n\phi}^i & \gamma_{n\Phi}^i \\
    \gamma'^i_{\epsilon\phi} & \gamma'^i_{n\phi} & \sigma_\phi & \sigma_\times \\
    \gamma'^i_{\epsilon\Phi} & \gamma'^i_{n\Phi} & \sigma'_\times & \sigma_\Phi
  \end{pmatrix}
  \begin{pmatrix}
    \dow_j T \\
    T\dow_j \frac{\mu}{T} - E_i \\
    \xi_t - \mu \\
    \ell(\Xi_t -\mu)
  \end{pmatrix}.
\end{equation}
The objects appearing in the matrix here have to be constructed out of the
zero-gradient structures $\delta^{ij}$, $\epsilon^{ij\ldots}$, and $\xi^i$,
supplemented with coefficients that are arbitrary functions of $T$, $\mu$,
$\dow^i\phi\dow_i\phi$, $\psi$. If we were only interested in the terms that
contribute linearly to the constitutive relations, we can ignore any dependence
on $\xi^i$, $\dow^i\phi\dow_i\phi$, and $\psi$. Further imposing
parity-symmetry, we have the allowed coefficients
\begin{gather}
  \kappa^{ij} = \kappa\delta^{ij},\quad
  \sigma_n^{ij} = \sigma_n\delta^{ij},\quad
  \gamma^{ij} = \gamma\delta^{ij},\quad
  \gamma'^{ij} = \gamma'\delta^{ij}, \nn\\
  \sigma_\phi, \quad
  \sigma_\Phi, \quad
  \sigma_\times, \quad
  \sigma'_\times,
\end{gather}
while all vector coefficients vanish. All coefficients are functions of $T$ and
$\mu$. Onsager's reciprocity relations~\cite{Onsager1,Onsager2} further impose
$\gamma'=\gamma$ and $\sigma'_\times = \sigma_\times$. The second law results in
the inequality relations
\begin{equation}
  \kappa,\sigma_\phi \geq 0, \qquad
  \sigma_n \geq \gamma^2/\kappa, \qquad
  \sigma_\Phi \geq \sigma_\times^2/\sigma_\phi.
\end{equation}

In addition, we have the ``hydrostatic sector'', characterised by corrections to the
entropy density
\begin{equation}
  \mathcal{S} = f_1\dow_i\xi^i
  - \frac{\ell\bar f_s}{T} \xi^i\Xi_i.
  \label{eq:calS-U1}
\end{equation}
The factor of $\ell$ in front of $\bar f_s$ is necessary because all dependence
on $\Phi$ must be expressible as a combination involving $\psi$. Indeed
$\ell\Xi_i = \ell\xi_i - \dow_i\psi$. The coefficient $\bar f_s$ characterises the response
of the system due to a background superfluid velocity due to the presence of $\Phi$.
If we only focus on linear corrections to
the constitutive relations, \eqref{eq:adiabaticity-U1} means that we only need
to consider the entropy density up to quadratic order in fields. Therefore,
without loss of generality, we can take the coefficient $\bar f_s$ to be
constant, while $f_1$ can be taken to be a linear function of $\epsilon$ and
$n$. Plugging \eqref{eq:calS-U1} into \eqref{eq:adiabaticity-U1}, we derive the
hydrostatic constitutive relations
\begin{align}
  \mathcal{E}^i_{\text{hs}}
  &= - T\mu\dow_i f_1
    - \ell \mu\bar f_s (\xi^i+\Xi^i) \nn\\
  \mathcal{J}^i_{\text{hs}}
  &= - T\dow_i f_1
    - \ell\bar f_s (\xi^i+\Xi^i) \nn\\
  \mathcal{K}_{\text{hs}}
  &= T\dow_i\dow^i f_1
    + T\sigma_\phi \lb
    \mu\frac{\dow f_1}{\dow \epsilon} 
    + \frac{\dow f_1}{\dow n} \rb \dow_i\xi^i
    + \ell\bar f_s \dow_i\Xi^i, \nn\\
  \mathcal{L}_{\text{hs}}
  &= \bar f_s \dow_i\xi^i,
\end{align}
along with
\begin{align}
  \mathcal{S}^i
  &= - f_1\dow_t\xi^i
    + \lb \dow_i f_1 + \frac{\ell\bar f_s}{T} \Xi_i \rb (\xi_t-\mu) \nn\\
  &\qquad
    + \frac{\ell\bar f_s}{T} \xi_i (\Xi_t-\mu),
\end{align}
where we have used the first-order equations of motion
\begin{equation}
  \dow_t\epsilon = \mu\sigma_\phi(\xi_t-\mu), \qquad
  \dow_t n = \sigma_\phi (\xi_t-\mu).
\end{equation}
Demanding $\mathcal{S}$ to be invariant under time-reversal symmetry, the
coefficient $f_1$ is not allowed. Additionally, it can be checked that the
coefficient $\bar f_s$ does not contribute to the linearised equations of
motion, when coupled to a homogeneous background $\Phi = \mu_0 t$. Note also
that, linearly, these coefficients can be removed by a redefinition of the
pseudo-Goldstone field $\phi \to \phi + (Tf_1 + \bar f_s\psi)/f_s$ and are only
physical if one has an unambiguous macroscopic notion of the pseudo-Goldstone
field. For these reasons, we have not considered these coefficients in the
remainder of our discussion.


\vspace{1em}
\noindent 
\emph{Modes}.---Focusing on isothermal fluctuations and employing the
definitions in \eqref{eq:U1-constraints}, we can read obtain the damped sound
modes
\begin{align}
  \omega
  &= \pm \sqrt{\omega_0^2+v_s^2k^2
    -\frac{1}{4} \lb\Gamma-\Omega+(D_n-D_\phi)k^2\rb^2} \nn\\
  &\qquad
    - \frac{i}{2}\lb k^2
    (D_n+D_\phi)+\Gamma+\Omega \rb.
\end{align}
Expanding this expression for $k^2\ll 1$, we find
\begin{align}
  \omega
  &= - \frac{i}{2}(\Gamma+\Omega)
    \pm \sqrt{\omega_0^2 -\frac{1}{4}(\Gamma-\Omega)^2}
    \nn\\
  &- \frac{i}{2} k^2 \lb D_n+D_\phi
    \pm i \frac{
    v_s^2-\frac{1}{2} (\Gamma-\Omega)(D_n-D_\phi)}
    {\sqrt{\omega_0^2 -\frac{1}{4}(\Gamma-\Omega)^2}} \rb.
\end{align}
The sound modes \eqref{eq:sound-U1} in the main text can be obtained from here
by ignoring the $\mathcal{O}(\ell^4,\ell^2k^2)$ corrections, i.e. assuming
$\omega_0\gg \Gamma,\Omega$.

\vspace{1em}
\noindent 
\emph{Correlation functions}.---We can also obtain the hydrodynamic predictions
for the retarded correlation functions. Specifically, at zero momentum we have
\begin{align}
  G^R_{nn}(\omega)
  &= \chi - \frac{\chi\omega(\omega +i\Omega)}
  {(\omega+i\Gamma)(\omega+i\Omega)-\omega_0^2}, \nn\\
  G^R_{\phi\phi}(\omega)
  &= \frac{\lambda^2}{\chi\omega_0^2}
    - \frac{
    \lambda^2\omega(\omega+i\Gamma)/(\chi \omega_0^2)}
    {(\omega+i\Gamma)(\omega+i\Omega)-\omega_0^2}, \nn\\
  G^R_{n\phi}(\omega)
  &= \frac{
    i \omega\lambda}
    {(\omega+i\Gamma)(\omega+i\Omega)-\omega_0^2},
    \nn\\
  G^R_{j^ij^j}(\omega)
  &=  - f_s\delta^{ij} + i \sigma_n\omega\delta^{ij},
    \label{eq:U1-correlators}
\end{align}
while all other correlators vanish. In general, these can be used to derive Kubo
formulas for the various transport coefficients.
In particular, it is easy to see that the Kubo formula \eqref{eq:lambdaK} for
$\lambda$ holds.

\section{Goldstone charge and $\lambda$ coefficient}

We defined the pseudo-Goldstone field $\phi$ to have unit charge under U(1)
transformations, i.e. $\phi\to\phi - \Lambda$. Normally, $\phi$ does not have
any well-defined macroscopic meaning and we can equally work with a rescaled
field $\tilde\phi = q\phi$ with charge $q$. As it turns out, by choosing the
charge $q=1/\lambda$, we can entirely remove $\lambda$ from the hydrodynamic
equations in the absence of background sources. Nonetheless, $\lambda$
resurfaces upon coupling the system to background electromagnetic fields and
thus has non-trivial effects on the correlation functions.

In order to see this explicitly, we consider the equations of motion following from
\eqref{eq:U1-constraints}, but turning on the gauge field $A_t$, $A_i$. Note
that we can always keep the background phase $\Phi=\mu_0t$ fixed by means
of a gauge transformation. Expressed in terms of $\tilde\phi$, we find 
\begin{align}
  \dow_t\delta\mu
  &= - \Gamma\delta\mu_A
    + D_n \lb \dow_i\dow^i\delta\mu_A + \dow_t \dow_i A^i \rb \nn\\
  &\qquad
    + q\lambda\frac{\tilde f_s}{\chi} \lb \dow_i\dow^i\delta\tilde\phi + q\dow_iA^i
    \rb
    - \frac{1}{q\lambda}\omega_0^2 \delta\tilde\phi, \nn\\
  \dow_t\delta\tilde\phi 
  &= q\lambda\delta\mu_A
    - \Omega\delta\tilde\phi
    + D_\phi \lb \dow_i\dow^i\delta\tilde\phi + q\dow_i A^i \rb,
    \label{eq:U1-linear-back}
\end{align}
where $\delta\mu_A = \delta\mu - A_t$. Note that $n = n_0+\chi\delta\mu$. We
have rescaled the superfluid density $\tilde f_s = f_s/q^2$. As we can clearly
see, upon setting $q=1/\lambda$, the coefficient $\lambda$ only appears coupled
to $A_i$.  This suggests that $\lambda$ should not appear in the $G^R_{nn}$,
$G^R_{\tilde\phi\tilde\phi}$, and $G^R_{n\tilde\phi}$ correlators, which is
indeed what we see from \eqref{eq:U1-correlators}. If one were only interested
in these correlators, $\lambda$ could be safely ignored.

However, $\lambda$ will still non-trivially affect the flux correlators due to
its coupling to $A_i$. In fact, it is easy to see from \eqref{eq:U1-correlators}
that the flux correlator now becomes
\begin{equation}
  G^R_{j^ij^j}(\omega)
  =  -\frac{\tilde f_s}{\lambda^2}\delta^{ij} + i \sigma_n\omega\delta^{ij}~~.
\end{equation}
This ensures that the Kubo formula \eqref{eq:lambdaK} for $\lambda$ remains
intact.

\section{Second law constraints in pinned viscoelastic hydrodynamics}

We now give details about the second law analysis for pinned viscoelastic
hydrodynamics. We introduce a gauge field $A_t$, $A_i$, which can be used to
compute the correlations of $n$, $\pi_i$ respectively. For technical simplicity,
we will omit introducing sources for $\tau^{ij}$, $\epsilon^t$, $\epsilon^i$,
which would require using Newton-Cartan geometry; see
e.g.~\cite{Jain:2020vgc}. The energy and momentum conservation equations take
the form
\begin{align}
  \dow_t\epsilon + \dow_i\epsilon^i
  &= E_i j^i - K_I\dow_t\phi^I - \ell L_I\dow_t\Phi^I, \nn\\
  \dow_t \pi^i + \dow_j\tau^{ij}
  &= E^i\rho + F^{ij} j_j
    + K_I\dow^i\phi^I + \ell L_I\dow^i\Phi^I, \nn\\
  \dow_t n + \dow_i \pi^i &= 0,
\end{align}
while the Josephson and continuity equations remains the same as
\eqref{eq:JoPinned}-\eqref{mom-conservation-crystal}. We parametrise the entropy
density to be
\begin{equation}
  s^t = s + \mathcal{S},
\end{equation}
where the thermodynamic entropy density $s$ is defined in
\eqref{eq:thermo-crystal} and $\mathcal{S}$ denotes gradient corrections. Using
the thermodynamic relations in \eqref{eq:thermo-crystal}, we can obtain
\begin{align}
  \dow_t s^t + \dow_i s^i
  &= - \frac{1}{T^2} \mathcal{E}^i\dow_i T
    - \frac{1}{T}\mathcal{T}^{ij} \dow_i u_j \nn\\
  &\hspace{-2em}
    - \frac{1}{T}\mathcal{K}_I\frac{\df\phi^I}{\df t}
    - \frac{\ell}{T}\mathcal{L}_I \frac{\df\Phi^I}{\df t}
    + \frac{\df}{\df t}\mathcal{S}
    + \dow_i\mathcal{S}^i,
    \label{eq:adiabaticity-crystal}
\end{align}
where we have identified the constitutive relations 
\begin{align}
  \epsilon^i
  &= (\epsilon-\mathcal{F})u^i
    + r_{IJ} \dow^i\phi^I \dow_t\phi^J
    + \mathcal{T}^{ij} u_j
    + \mathcal{E}^i, \nn\\
  \tau^{ij}
  &= \rho\, u^i u^j
    - \mathcal{F}\,\delta^{ij}
    - r_{IJ} \dow^i\phi^I \dow^j\phi^J
    + \mathcal{T}^{ij}, \nn\\
  K_I &= - \dow_i\lb r_{IJ}\dow^i\phi^J \rb - \ell m^2\psi_I
        + \mathcal{K}_I, \nn\\
  L_I &= m^2 \psi_I + \mathcal{L}_I, \nn\\
  s^i
  &= s^t u^i 
    + \frac{1}{T}\mathcal{E}^i
    + \mathcal{S}^i.
\end{align}

Similarly to the U(1) case, the right-hand side of \eqref{eq:adiabaticity-crystal}
is required to be a positive semi-definite quadratic form. This results in the
``non-hydrostatic'' constitutive relations
\begin{equation}
  \def\arraystretch{1.2}
  \begin{pmatrix}
    \frac1T\mathcal{E}^i \\
    \mathcal{T}^{ij} \\
    \mathcal{K}_I \\ \mathcal{L}_I 
  \end{pmatrix}
  = -
  \begin{pmatrix}
    \sigma_\epsilon^{ij} & \chi^{ikl}
    & -\gamma_{\phi}{}^{i}_J & -\gamma_{\Phi}{}^{i}_{J} \\
    \chi^{ijk} & \eta^{ijkl}
    & \chi_{\phi}{}^{ij}_{I} & \chi_{\Phi}{}^{ij}_{I} \\
    \gamma_{\phi}{}^{k}_{I} & \chi_{\phi}{}^{kl}_{I}
    & \sigma^\phi_{IJ} & \sigma^\times_{IJ} \\
    \gamma_{\Phi}{}^{k}_{I} & \chi_{\Phi}{}^{kl}_{I}
    & \sigma^\times_{IJ} & \sigma^\Phi_{IJ}
  \end{pmatrix}
  \begin{pmatrix}
    \dow_k T \\
    \dow_{(k} u_{l)} \\
    \frac{\df}{\df t}\phi^J \\
    \ell\frac{\df}{\df t}\Phi^J
  \end{pmatrix},
\end{equation}
where all the objects in the coefficient matrix have to be made out of
$\delta^{ij}$, $\epsilon^{ij\ldots}$, $\dow_i\Phi^I$, supplemented with
arbitrary transport coefficients that are functions of $T$, $\mu$, $h^{IJ}$,
$\psi^I$. We have already imposed the Onsager's reciprocity relations in the
matrix above. Assuming the crystal to be isotropic and parity-preserving, and
focusing only on the terms that contribute to the linearised constitutive
relations, we find
\begin{gather}
  \sigma_\epsilon^{ij} = \sigma_\epsilon\delta^{ij}, \quad
  \eta^{ijkl} = \eta(\delta^{ik}\delta^{jl}-\delta^{il}\delta^{jk})
  + \lb \zeta - {\textstyle\frac{2}{d}}\eta \rb \delta^{ij}\delta^{kl}, \nn\\
  \sigma^\phi_{IJ} = \sigma_\phi h_{IJ}, \quad
  \sigma^\Phi_{IJ} = \sigma_\Phi h_{IJ}, \nn\\
  \gamma_{\phi}{}^i_J = \gamma_{\phi} e^i_J, \quad
  \gamma_{\Phi}{}^i_J = \gamma_{\Phi} e^i_J, \quad
  \sigma^\times_{IJ} = \sigma_\times h_{IJ},
\end{gather}
and all others zero. Here $e^i_I = h_{IJ}\dow^i\phi^J$. All coefficients are
functions of $T$ and $\mu$. The second law results in a set of inequality
constraints on these coefficients
\begin{equation}
  \eta,\zeta,\sigma_\epsilon,\sigma_\phi\geq 0, \qquad
  \sigma_\Phi \geq \sigma_\times^2/\sigma_\phi.
\end{equation}

For the ``hydrostatic sector'', we need to consider the most general first order
gradient corrections in $\mathcal{S}$. It was already found
in~\cite{Armas:2019sbe,Armas:2020bmo} that, assuming the crystal to be
isotropic, there are no allowed terms in the absence of the background field
$\Phi^I$. However, in the presence of $\Phi^I$ we can include the term
\begin{equation}
  \mathcal{S} = \frac{1}{T}\bar f_{IJ}\gamma^{IJ},
\end{equation}
where
\begin{equation}
  \gamma^{IJ}
  = \half\lb 
  - 2\dow_k\phi^{(I} \dow^k\psi^{J)}
  + \ell h^{IJ} - \ell \delta^{IJ}\rb,
\end{equation}
is defined so that, linearly, we have
$\gamma^{IJ}\approx \ell \dow^{(I}\delta\Phi^{J)}$. We can further require that
$\bar f_{IJ}$ is at least linear in fluctuations, because the constant
contribution can be removed using a total derivative term
$\dow_i\delta\Phi^i$. We hence have
\begin{equation} \label{eq:barfcrystal}
  \bar f_{IJ} = \bar\alpha\,h_{IJ}
  + \bar C_{IJKL} u^{KL}.
\end{equation}
where $\bar\alpha(T_0,\mu_0)=0$. This coefficient can be understood
as the response of the crystal to a background strain due to the presence of the background lattice. 
Ignoring non-linear terms in the constitutive
relations, we have 
\begin{align}
  \mathcal{E}^i_{\text{hs}}
  &= 0, \nn\\
  \mathcal{T}^{ij}_{\text{hs}}
  &= - \lb \ell\bar f_{IJ}
    - \gamma^{KL} \bar C_{KLIJ} \rb \dow^i\phi^I\dow^j\phi^J \nn\\
  &\qquad
    - \lb (Ts+\mu\rho)
    \frac{\dow\bar\alpha}{\dow \epsilon}
    + \rho
    \frac{\dow\bar\alpha}{\dow\rho} \rb h_{IJ}\gamma^{IJ} \delta^{ij}, \nn\\
  \mathcal{K}_I^{\text{hs}}
  &= \dow_i \lb \bar C_{KLIJ} \gamma^{KL} \dow^i\phi^J\rb, \nn\\
  \mathcal{L}_I^{\text{hs}}
  &= -\dow_i \lb \bar f_{IJ} \dow^i\phi^{J} \rb,
\end{align}
along with
\begin{align}
  \mathcal{S}^i
  &= \frac{1}{T}\bar f_{IJ}
    \lb \dow^i\psi^{J}  \frac{\df \phi^{I}}{\df t}
    - \ell \dow^i\phi^J \frac{\df\Phi^I}{\df t} \rb \nn\\
  &\qquad
    - \frac{1}{T} \bar C_{KLIJ} \gamma^{KL} \dow^i\phi^J
     \frac{\df}{\df t}\phi^I.
\end{align}
We can remove one component of $\bar C_{IJKL}$ using the redefinition of
pseudo-Goldstone fields $\phi^I\to \phi^I + a\,\psi^I$.


\section{Linear pinned viscoelastic crystals}

Linearising the equations on a homogeneous background $\Phi^I=x^I$, we can
obtain the Josephson equation
\begin{align}
  \dow_t\delta\phi^i
  &= \lambda u^i - \Omega\delta\phi^i
    + \gamma_m \dow^i\mu
    + \gamma_T \dow^iT \nn\\
  &\qquad
    + 2 D_\phi^\perp \dow_j\dow^{[j}\delta\phi^{i]}
    + D_\phi^\| \dow^i\dow_k\delta\phi^k,
\end{align}
where
\begin{gather}
  \gamma_m = - \frac{B \alpha_m}{\sigma_\phi}, \qquad
  \gamma_T = \frac{\gamma_{\phi}}{\sigma_\phi}
  - \frac{B \alpha_T}{\sigma_\phi}, \nn\\
  \alpha_m = -d \frac{\dow\ln\alpha}{\dow \mu}, \qquad
  \alpha_T = -d \frac{\dow\ln\alpha}{\dow T}.
\end{gather}
Here $\alpha_m$ is the mass expansion coefficient, while $\alpha_T$ is the
thermal expansion coefficient. For the conservation equations, we find
\begin{align}
  \epsilon^i
  &= \lb \epsilon+p + T\lambda_T\rb u^i
    + T\Omega_s \delta\phi^i 
    - \kappa_m \dow^i\mu
    - \kappa\, \dow^iT \nn\\
  &\qquad
    - T \gamma_{\phi} \lb D_\phi^\| \dow^i\dow_k\delta\phi^k
    + 2D_\phi^\perp \dow_j \dow^{[j}\delta\phi^{i]} \rb, \nn\\[0.5em]
  \tau^{ij}
  &= p_m \delta^{ij}
    - 2\lambda G\, \dow^{\langle i}\delta\phi^{j\rangle}
    - \lambda B\, \delta^{ij} \dow_k\delta\phi^k \nn\\
  &\qquad
    - 2\eta\, \dow^{\langle i} u^{j\rangle}
    - \zeta\, \dow_k u^k \delta^{ij}
    - \ell\mathcal{X}^{ij}, \nn\\[0.5em]
  \ell L^i
  &= - \rho\lambda \omega_0^2 \delta\phi^i
    - \Gamma \pi^i
    - \lambda_T \dow^i T
    - \ell\dow_j\mathcal{X}^{ij},
\end{align}
where we have further defined the mechanical pressure $p_m$, mass conductivity
$\kappa_m$, thermal conductivity $\kappa$, heat damping coefficient $\Omega_s$,
and a new coefficient $\lambda_T$ as
\begin{gather}
  p_m = p - \lambda Bd\,\delta\ln\alpha, \nn\\
  \kappa_m = - \frac{T\gamma_{\phi}  B \alpha_\mu}{\sigma_\phi}, \quad 
  \kappa = T\sigma_\epsilon + \frac{T\gamma_{\phi}^2}{\sigma_\phi}
  - \frac{T\gamma_{\phi}  B \alpha_T}{\sigma_\phi}, \nn\\
  \Omega_s = \frac{\gamma_{\phi}\ell^2 m^2}{\sigma_\phi}, \quad
  \lambda_T = \ell\lb \gamma_{\Phi}
  - \frac{\sigma_\times}{\sigma_\phi} \gamma_{\phi} \rb,
\end{gather}
as well as 
\begin{align}
  \mathcal{X}^{ij}
  &=
    2\lb \bar G -\frac{\sigma_\times}{\sigma_\phi}G \rb
    \dow^{\langle i}\delta\phi^{j\rangle}
    + \lb \bar B -\frac{\sigma_\times}{\sigma_\phi}B \rb \delta^{ij}
    \dow_k\delta\phi^k \nn\\
  &\qquad
    + \delta_{ij} \lb \lb \bar B 
      -\frac{\sigma_\times}{\sigma_\phi} B \rb d \delta\alpha
    + \delta\bar\alpha \rb.
    \label{eq:Xij}
\end{align}
Note that $\mathcal{X}^{ij}$ identically drops out from the equations of motion,
and is the only contribution that contains $\bar f_{IJ}$. However,
$\mathcal{X}^{ij}$ still non-trivially affects the stress tensor and respective
correlation functions. For the record, let us also note the heat/entropy flux
\begin{align}
  s^i
  &= \lb s + \lambda_T\rb u^i
    + \Omega_s \delta\phi^i 
    - \frac{\kappa_m}{T} \dow^i\mu
    - \frac{1}{T}\kappa\, \dow^iT \nn\\
  &\qquad
    - \gamma_{\phi} \lb D_\phi^\| \dow^i\dow_k\delta\phi^k
    + 2D_\phi^\perp \dow_j \dow^{[j}\delta\phi^{i]} \rb.
\end{align}

From here, we derive another damping-attentuation relation in
energy/entropy/heat flux
\begin{equation}
  \Omega_s = \gamma_{\phi} D_\phi^\perp k_0^2,
\end{equation}
where $k_0^2 = \ell^2 m^2/G$. This relation recently appeared
in~\cite{Delacretaz:2021qqu}, where the authors derived it using the locality of
hydrodynamic constitutive relations. We also find a new coefficient $\lambda_T$
that modifies the energy and entropy flux at thermodynamic level, and
contributes to sourcing momenta.

\section{Mode spectrum of pinned viscoelastic crystals}

We can use the linearised equations of motion to derive the mode spectrum of
pinned viscoelastic hydrodynamics. In the transverse sector, we find a phonon
sound mode with the dispersion relation similar to the U(1) case, namely
\begin{align}
  \omega
  &= \pm \sqrt{\omega_0^2+v_\perp^2k^2} - \frac{i}{2}\lb k^2
    (D_\pi^\perp+D_\phi^\perp)+\Gamma+\Omega \rb,
\end{align}
where
\begin{equation}
  v_\perp^2 = \frac{\lambda^2G}{\rho}, \qquad
  D_\pi^\perp = \frac{\eta}{\rho}.
\end{equation}
The longitudinal sector is considerably more involved. Focusing on isothermal
configurations, we find a damped sound mode and a crystal diffusion mode
\begin{align}
  \omega
  &= \pm\sqrt{\omega_0^2+v_\|^2k^2} \nn\\
  &\qquad
    -\frac{i}{2} \lb
    D_{s}^\| k^2
    - \frac{\rho_m^2}{v_\|^2\chi^2}\frac{\Omega k^2}{\omega_{0}^2+v_\|^2k^2}
    + \Gamma + \Omega
    \rb, \nn\\
  \omega
  &= -\frac{i k^2 \rho/\chi}{\omega_0^2+v_\|^2k^2}
    \lb D_\phi^\| k^2+\Omega \rb,
\end{align}
where we have defined
\begin{align}
  v_\|^2
  &= \frac{\rho_m^2/\chi + \lambda^2(B+2\frac{d-1}{d}G)}{\rho}, \nn\\
  D_s^\|
  &= \frac{\rho(v_\|^2 -\rho_m/\chi)^2}{\sigma_\phi  v_\|^2}
    + \frac{\zeta + 2{\textstyle\frac{d-1}{d}}\eta}{\rho},
\end{align}
along with mechanical mass density $\rho_m = \rho + \lambda B \alpha_\mu$ and
susceptibility $\chi = \dow\rho/\dow\mu$. We have taken
$\ell\sim\mathcal{O}(\dow)$ in the expressions above. We again note that
$\lambda$ non-trivially affects the various speeds of mode propagation. While
solving the linearised equations, it is useful to note that in the isothermal
limit, $\chi\delta\mu = \delta\rho - B\alpha_\mu \dow_k\delta\phi^k$.

Turning back on the background fields, we can compute the following correlation
functions at zero momentum
\begin{align}
  G^R_{\pi^i\pi^j}(\omega,k=0)
  &= \rho\,\delta^{ij} - \frac{\rho\,\omega(\omega +i\Omega)\delta^{ij}}
  {(\omega+i\Gamma)(\omega+i\Omega)-\omega_0^2}, \nn\\
  G^R_{\phi^i\phi^j}(\omega,k=0)
  &= \frac{\delta^{ij}\lambda^2}{\rho\omega_0^2}
    - \frac{\lambda^2
    \omega(\omega+i\Gamma)/(\rho\omega_0^2)\delta^{ij}}
    {(\omega+i\Gamma)(\omega+i\Omega)-\omega_0^2}, \nn\\
  G_{\pi^i\phi^j}^R(\omega,k=0)
  &=\frac{-i\omega\lambda \delta^{ij}}{
    (\omega+i\Gamma)(\omega+i\Omega)
    - \omega_0^2}~.
\end{align}
Computing $\tau^{ij}$ correlators is beyond the scope of this work and would
require coupling the system to curved space. Note that the coefficients \
\eqref{eq:barfcrystal} do not affect these three correlators above.

\section{Comparison with previous works} 

In the previous version of our paper, we pointed out certain discrepancies with
the work of~\cite{Delacretaz:2021qqu}. These have now been resolved by the
authors of \cite{Delacretaz:2021qqu} in an updated version of their paper; 
we present a detailed comparison below.

Let us start with the U(1) model. Our constitutive relations in
\eqref{eq:U1-constraints} trivially reduce to those in~\cite{Delacretaz:2021qqu}
upon setting $\sigma_\times=\bar f_s=0$ (i.e. $\lambda=1$) and matching the
conventions $\phi\to-\phi$, $A_t\to-A_t$, $A_i\to-A_i$. Our results also match
with~\cite{Ammon:2021slb} in this limit upon matching the conventions
$\phi\to-\phi$. In an updated version of their paper, the authors
of~\cite{Delacretaz:2021qqu} verified that their formalism does allow for
nonzero $\sigma_\times$ and $\bar f_s$ in the presence of background gauge
fields. We find the new mapping between various coefficients
\begin{gather}
  \hat\chi_{nn} = \chi, \qquad
  \hat c_s^2 = \frac{f_s + 2\ell\bar f_s}{\chi}, \qquad
  \hat\omega_0^2 = \frac{\ell^2m^2}{\chi}, \nn\\
  \hat D_n = \frac{\sigma}{\chi}, \qquad
  \hat D_\phi = \frac{(1-\ell\sigma_\times/\sigma_\phi)(f_s + 2\ell\bar f_s)}{\sigma_\phi},\nn\\
  \hat\Gamma
  = \frac{\ell^2}{\chi}\lb \sigma_\Phi - \frac{\sigma_\times^2}{\sigma_\phi} \rb, \nn\\
  \hat\kappa =
  \frac{-\ell\bar f_s}{  f_s + 2\ell\bar f_s}, \qquad
  \hat\sigma = \frac{\ell\sigma_\times}{\sigma_\phi} = \lambda-1.
\end{gather}
For clarity, we have denoted all the coefficients in~\cite{Delacretaz:2021qqu}
with a hat.





To compare our results with~\cite{Delacretaz:2021qqu} in the pinned crystal
case, we need to perform the following transformations to the constitutive
relations
\begin{align}
  \tau^{ij}
  &\to \tau^{ij} + \ell\mathcal{X}^{ij}
  + 2\lb \dow^j\delta\phi^i - \delta^{ij}\dow_k\delta\phi^k \rb, \nn\\
  L^i
  &\to L^i + \dow_j\mathcal{X}^{ij},
\end{align}
which leave the equations of motion invariant at the linearised
level. $\mathcal{X}^{ij}$ was defined in \eqref{eq:Xij}. Note, however, that the
transformed quantities should not be used to reliably predict the hydrodynamic
correlation functions involving stress. Focusing on $d=2$ spatial dimensions,
this results in
\begin{align}
  s^i
  &= \lb s + \lambda_T\rb u^i
    + \Omega_s \delta\phi^i 
    - \frac{\kappa_m}{T} \dow^i\mu
    - \frac{\kappa}{T} \dow^iT \nn\\
  &\qquad
    - \gamma_{\phi} D_\phi^\| \dow^i\dow_k\delta\phi^k
    - 2\gamma_{\phi} D_\phi^\perp \dow_j
    \dow^{[j}\delta\phi^{i]}, \nn\\[0.5em]
  \tau^{ij}
  &= p_m \delta^{ij}
    - 2\lambda G\, \dow^{[i}\delta\phi^{j]}
    - \lambda (B + G) \delta^{ij} \dow_k\delta\phi^k \nn\\
  &\qquad
    - 2\eta\, \dow^{\langle i} u^{j\rangle}
    - \zeta\, \dow_k u^k \delta^{ij}, \nn\\[0.5em]
  \dow_t\delta\phi^i
  &= \lambda u^i - \Omega\delta\phi^i
    + \gamma_m \dow^i\mu
    + \gamma_T \dow^iT \nn\\
  &\qquad
    + 2 D_\phi^\perp \dow_j\dow^{[j}\delta\phi^{i]}
    + D_\phi^\| \dow^i\dow_k\delta\phi^k.
    \label{eq:transformed-consti}
\end{align}
These should be compared to~\cite{Delacretaz:2021qqu} in the Galilean setting,
i.e. upon setting $j^i = \pi^i$. Note that the displacement field $u^i$
of~\cite{Delacretaz:2021qqu} is identified with our $\delta\phi^i$, their fluid
velocity $v^i$ is our $u^i$, their heat current $j_Q^i$ is our $Ts^i$. Firstly,
the Galilean constraint implies for their transport coefficients
\begin{gather}
  \hat n = \hat\chi_{\pi\pi}, \qquad
  \hat\gamma_{3c} = - \hat\chi_{n\lambda_\|}\hat\xi(B+G), \nn\\
  \hat\Omega_n = \hat\sigma_0 = \hat\alpha_0 = \hat{\bar\alpha}_0
  = \hat\gamma_{1l} = 0.
\end{gather}
We have again used a hat for the coefficients in~\cite{Delacretaz:2021qqu} to
avoid confusion. The mapping between the remaining coefficients follows as
\begin{gather}
  \hat p = p_m, \qquad
  \hat\chi_{\pi\pi} = \rho, \nn\\
  \hat\chi_{n\lambda_\|} = \frac{B\alpha_\mu}{B+G}, \qquad 
  \hat\chi_{s\lambda_\|} = \frac{B \alpha_T}{B+G}, \nn\\
  \hat{\bar\kappa}_0= \kappa + \frac{T\gamma_\phi B\alpha_T}{\sigma_\phi}, \quad
  \hat\gamma_{2l} = \frac{\gamma_\phi}{\sigma_\phi}, \nn\\
  \hat\xi = \frac{1}{\sigma_\phi}, \quad
  \hat\gamma_{3h} = \gamma_T.
\end{gather}
With these identifications, we find that the results
of~\cite{Delacretaz:2021qqu} exactly match our \eqref{eq:transformed-consti},
modulo the new coefficients $\lambda$ and $\lambda_T$ due to explicit symmetry
breaking.

\clearpage

\end{document}